\documentclass[a4paper,12pt]{article}

\usepackage[T2A]{fontenc}
\usepackage[utf8]{inputenc}
\usepackage[english]{babel}
\usepackage{amsmath}
\usepackage{amsfonts}
\usepackage{mathrsfs}
\usepackage{amssymb}
\usepackage{cite}

\numberwithin{equation}{section}

\newcommand{\be}{\begin{equation}}
\newcommand{\ee}{\end{equation}}
\newcommand{\bea}{\begin{eqnarray}}
\newcommand{\eea}{\end{eqnarray}}

\newcommand{\p}[1]{(\ref{#1})}

\begin{document}

\begin{titlepage}

\vspace*{0.7cm}

\begin{center}
{\LARGE\bf Lagrangian formulation }

\vspace{0.4cm}

{\LARGE\bf for free $6D$ infinite spin field}

\vspace{1.2cm}

{\large\bf I.L.\,Buchbinder$^{1,2,3}$\!\!,\,\,
S.A.\,Fedoruk$^1$\!,\,\,  A.P.\,Isaev$^{1,5}$\!\!,\,\,
V.A.\,Krykhtin$^{2,4}$}

\vspace{10mm}

\ $^1${\it Bogoliubov Laboratory of Theoretical Physics,
Joint Institute for Nuclear Research, \\
141980 Dubna, Moscow Region, Russia}, \\
{\tt buchbinder@theor.jinr.ru, fedoruk@theor.jinr.ru,
isaevap@theor.jinr.ru}

\vskip 0.4cm

\ $^2${\it Department of Theoretical Physics,
Tomsk State Pedagogical University, \\
634041 Tomsk, Russia}, \\
{\tt joseph@tspu.edu.ru, krykhtin@tspu.edu.ru}

\vskip 0.4cm

\ $^3${Tomsk University of Control Systems and Radioelectronics
(TUSUR), \\634034, Tomsk, Russia}

\vskip 0.4cm

\ $^4${\it National Research Tomsk State  University,}\\{\em Lenin
Av.\ 36, 634050 Tomsk, Russia}

\vskip 0.4cm

\ $^5${\it Faculty of Physics, Lomonosov Moscow State University,
119991 Moscow, Russia}

\end{center}

\vspace{8mm}

\begin{abstract}
We construct a Lagrangian that describes the dynamics of a
six-dimensional free infinite (continuous) spin field in $6D$
Minkowski space. The Lagrangian is formulated in the framework of
the BRST approach to higher spin field theory and is based on a
system of constraints defining an irreducible representation of the
corresponding Poincar\'e group. The field realization of generators
in the $6D$ Poincar\'e algebra and the second-, fourth-, and
sixth-order Casimir operators are obtained in explicit form using
additional spinor coordinates. Specific aspects of such a
realization in six dimensions are discussed. We derive the
conditions that determine the irreducible representation $6D$
infinite spin field and reformulate them as operators in the Fock
space forming a first-class algebra in terms of commutators. These
operators are used to construct the BRST charge and the
corresponding Lagrangian. We prove that the conditions of the
irreducible representation are reproduced as the consequence of the
Lagrangian equations of motion, which finally provides the
correctness of the results obtained.
\end{abstract}

\vspace{10mm}

%\bigskip

%\phantom{Keywords: }

%\vspace{1cm}

\end{titlepage}

\setcounter{footnote}{0}
\setcounter{equation}{0}

\newpage

\setcounter{equation}0
\section{Introduction}

In this paper, we continue our work on the study of massless
irreducible field representations of the six-dimensional Poincare
group \cite{BFIP,BFI-21,BFI-22}. Here we focus on aspects of the
Lagrangian formulation for a free field that carries out an
irreducible representation with infinite spin. Modern interest in
multidimensional field theories is largely due to the fact that all
of them can be treated as low-energy approximations of superstring
theory, which is naturally formulated in ten dimensions. However, in
each concrete dimension, field models have specific features
that assume independent study. Among these features, a special place
should be given to aspects of symmetry, primarily space-time
symmetry, which underlie the classification of particles. This
explains a certain interest in the study of field representations of
the Poincar\'e group in higher dimensions (see e.g. \cite{W},
\cite{KUZ}, \cite{BFIP}, \cite{BFI-21} and the references therein).

Massive and massless irreducible representations of the
four-dimensional Poincar\'e group were constructed in the seminal
works of Wigner \cite{Wigner39,Wigner47} and Bargmann and Wigner
\cite{BargWigner} and are extremely widely used in field theory and
mathematical physics in general. The algebraic description of the
Poincar\'e group in any dimensions is well known (see e.g. the books
\cite{Isaev1,Isaev2}), including field representations \cite{BB1},
\cite{BB2}.

Recently, there has been a surge of interest in constructing
Lagrangian models for a special class of irreducible massless
representations of the Poincar\'{e} group called representations of
infinite (continuous) spin\footnote{Two names are used in the
literature for such representations: either infinite spin
representations or continuous spin representations. We will use the
name of the infinite spin representation.} (see, e.g., the review
\cite{BekSk} and earlier references therein, and recent papers
\cite{BekMou,Bekaert:2017xin,Najafizadeh:2017tin,HabZin,AlkGr,Metsaev18,BFIR,
BuchKrTak,R,BuchIFKr,ACG18,Metsaev18a,BFI,Metsaev19,BKSZ,MN20,MN22,BuchIFKr22,STZ}).
Such representations have the remarkable properties that their
irreducible multiplets contain an infinite number of helicities,
which makes them analogous to multiplets in string theory.
Therefore, one can expect that infinite spin field models may
possess properties similar to string models. This circumstance
emphasizes the importance of studying the classical and quantum
structure of infinite spin field theories. Taking into account that
string models lead to field theories in higher dimensions, it would
be natural to study  infinite spin filed theories  in various
dimensions.

In this paper we develop Lagrangian formulations for the bosonic
field carrying out the $6D$ irreducible representation of the
Poincar\'e group. As we have already noted, although the algebraic
description of finite dimensional representations of the Poincar\'e group is well known
for any space-time dimensions, the description of unitary field
representations in different dimensions has certain, sometimes
significant, differences. First of all, this concerns the explicit
expressions for the generators of the Poincar\'e group and the
independent Casimir operators, the number of which increases with
the space-time dimension. In turn, this leads to specific relations
for fields that define an irreducible infinite spin representation
in each dimension. As we will see, such relations in the $6D$ case
have a very special form.

The paper is organized as follows. In section 2, we describe the
irreducible infinite spin representations of the Poincar\'e group in
six-dimensional Minkowski space in suitable form for our aims and
derive the conditions that should be imposed on the field to get an
irreducible representation. Section 3 is devoted to the derivation of the
Lagrangian in the framework of the BRST approach to higher spin field
theories (see e.g., \cite{BKL,BKS} and the references therein).
First of all, we reformulate the conditions defining the $6D$
infinite spin irreducible field as operator constraints in the
Fock space and construct on their basis a BRST charge acting in
this space. The use of the BRST charge allows us to derive the Lagrangian
and gauge transformations for the model under consideration
in terms of both Fock space vectors and conventional
spin-tensor fields. We prove that the equations of motion for the
obtained Lagrangian reproduce the conditions defining the infinite
spin irreducible representation for fields.

\setcounter{equation}0
\section{$6D$ infinite spin representations}

\subsection{Representation in the space with spinorial operator}

Let us start with a brief survey of massless  irreducible
representations of the six-dimensional Poincar\'{e} group. For details
of such a construction see ref. \cite{BFI-22} (see also \cite{BFIP,BFI-21}).

Consider these representations in the linear space $V$ of vectors $\Psi$ and
assume that the space-time and momentum coordinates $x^{\mu},\, p_{\mu}$ and spinor
operators $\xi_\alpha^i,\, \rho^{\alpha i}$ act in this space. The
commutation relations for these operators have the form:
\begin{equation}
\label{op-v} [x^{\mu},p_{\nu}]= i\delta^{\mu}_{\nu}\,,
\end{equation}
\begin{equation}
\label{op-s} \left[ \xi_\alpha^i\,, \rho^\beta_j\right] =
i\delta_\alpha^\beta \delta^i{}_j\,,
\end{equation}
the other commutators are equal to zero.
The Hermitian operators of the coordinate $x^{\mu}=(x^{\mu})^\dagger$ and momentum
$p_{\mu}=(p_{\mu})^\dagger$ are the components of the
six-vectors, $\mu=0,1,\ldots,5$. The operators $\xi_\alpha^i$,
$\rho^{\alpha i}$ are the $\mathrm{SU}(2)$ Majorana-Weyl
spinors\footnote {We use the spinor conventions of the works
\cite{BFIP,BFI-21,BFI-22}.}, where $\alpha=1,2,3,4$ and $i=1,2$ are
the spinorial $\mathrm{SU}^*(4)$ and internal $\mathrm{SU}(2)$
indices, respectively. The Hermitian conjugation for the spinor
operators is defined as follows:
\begin{equation}
\label{xi} (\xi_\alpha^i)^\dagger
=\epsilon_{ij}B_{\dot\alpha}{}^\beta \xi_\beta^j\,,\qquad
(\rho^{\alpha i})^\dagger = \epsilon_{ij}\rho^{\beta
j}(B^{-1})_\beta{}^{\dot\alpha}\,,
\end{equation}
where $B_{\dot\alpha}{}^\beta$ is the matrix of the complex
conjugation, and the antisymmetric tensors $\epsilon_{ij}$,
$\epsilon^{ij}$ are fixed by the conditions
$\epsilon_{12}=\epsilon^{21}=1$ (see \cite{BFI-21} for details).

As usual, the operators $p_\mu$ generate space-time
translations. In this case, the generators $\{P_\mu, M_{\mu\nu} \}$
of the Poincar\'e algebra $\mathfrak{iso}(1,5)$ are realized in $V$ as
\begin{equation}
\label{PM-gen} P_\mu=  p_\mu \,, \qquad M_{\mu\nu}=
x_{\mu}p_{\nu}-x_{\nu}p_{\mu}+S_{\mu\nu} \,,
\end{equation}
where the spin part of the Lorentz group generators looks like
\begin{equation}
\label{M-1} S_{\mu\nu} = \xi_\alpha^i
(\tilde\sigma_{\mu\nu})^{\alpha}{}_{\beta}\rho^\beta_i \,.
\end{equation}

The massless representation subspace in $V$ is distinguished by the
following condition for the quadratic Casimir operator $C_2 = P^2 =
P^{\mu}P_{\mu}$ of $\mathfrak{iso}(1,5)$:
\begin{equation}
\label{P2-0}
p^{\mu}p_{\mu}\,\Psi =  0\,.
\end{equation}
In this subspace, i.e. when the relation \p{P2-0} is satisfied, the
fourth-order Casimir operator takes the following form
\cite{BFI-22}:
\begin{equation}
\label{C4-expr} C_4= - \,\tilde\ell\,\ell  \,,
\end{equation}
where the scalar operators $\ell$, $\tilde\ell$ are defined as
\begin{equation}
\label{l-expr}
\ell \ := \ \frac12\, \rho^\alpha_i
p_{\alpha\beta}\rho^\beta{}^i\,,\qquad\quad \tilde\ell \ := \
\frac12\,  \xi_\alpha^i \tilde p^{\alpha\beta}\xi_\beta{}_i \,,
\end{equation}
where $p_{\alpha\beta}:=p_\mu(\sigma^{\mu})_{\alpha\beta}$,
$\tilde p^{\alpha\beta}:=p_\mu(\tilde\sigma^{\mu})^{\alpha\beta}$,
and the matrices $\sigma^\mu$ and $\tilde\sigma^\mu$ are skew symmetric
(see Appendix\,B in  \cite{BFI-21}).
The algebra of operators \p{l-expr} is written in the form
\begin{equation}
\label{l-com}
[\tilde\ell,\ell]=N\, p^\mu p_\mu\,,\qquad
N:=\frac{i}{2}\, \{\xi_\alpha^i,\rho^\alpha_i\} \,,
\end{equation}
and $\{.,. \}$ is anticommutator.
Besides, the operators \p{l-expr} commute with the Poincar\'e group
generators \p{PM-gen}:
\begin{equation}
\label{l-PM} [P_\mu,\ell]=[P_\mu,\tilde\ell]=0\,\qquad
[M_{\mu\nu},\ell]=[M_{\mu\nu},\tilde\ell]=0\,.
\end{equation}

The infinite spin representation is characterized by the condition
that the fourth-order Casimir operator has nonzero negative
eigenvalue
\begin{equation}
\label{C4-const} C_4\,\Psi \ = \ -\,\mu^2\,\Psi \,,
\end{equation}
where $\mu \neq 0$ is the dimensional real parameter which can be
taken positive $\mu \in \mathbb{R}_{> 0}$ without loss of
generality. Using relations \p{C4-const} and \p{C4-expr}, one can
see that it is sufficient to define infinite spin states by the
constraints
\begin{equation}
\label{C4-constr}
\ell\,\Psi \ = \ \mu\,\Psi \,,\qquad
\tilde\ell\,\Psi \ = \ \mu\,\Psi \, .
\end{equation}
Note that we consider here, as in \cite{BFI-22}, the dimensionless
spinor operators $\xi_\alpha^i$, $\rho^{\alpha i}$, unlike the
dimensional twistor operators  in \cite{BFI-21}.

For massless representations \p{P2-0} in the infinite spin case
\p{C4-const}, the sixth-order Casimir operator \cite{BFIP} is
written as
\begin{equation}
\label{C6-Cas}
C_6\,|\Psi\rangle \ = \ -\, \mu^2\,
J_{a}J_{a}\,|\Psi\rangle\,,
\end{equation}
that contains the operators
\begin{equation}\label{Ta-def}
J_{a} \ := \ \frac{i}{2}\,\xi_\alpha^i(\sigma_{a})_i{}^j
\rho^\alpha_j\,,
\end{equation}
where $\sigma_{a}$, $a=1,2,3$ are the Pauli matrices. The operators
$J_{a}$ form the $\mathfrak{su}(2)$ algebra
\begin{equation}\label{Ta-alg}
[J_{a},J_{b}]= i\epsilon_{abc}J_{c}\,.
\end{equation}

As it was shown in \cite{BFIP}, on the states of the irreducible
infinite spin representation the operator $C_6$ takes the values
\begin{equation}
\label{C6-ir} C_6\,|\Psi\rangle \ = \ -\,\mu^2\,
s(s+1)\,|\Psi\rangle\,,
\end{equation}
where $s$ is a nonzero integer or half-integer number, $s \in \mathbb{Z}_{\geq 0}/2$.
Therefore, the states corresponding to the
infinite spin irreducible representation obey the constraints
\begin{equation}\label{eq-3}
J_{a}J_{a}\,|\Psi\rangle \ = \ s(s+1)\,|\Psi\rangle \,,
\end{equation}
where the operators  $J_{a}$ are defined in \p{Ta-def}.

Note that the $\mathfrak{su}(2)$ algebra generators \p{Ta-def}
commute with the $\mathrm{SU}(2)$ scalar operators \p{l-expr}:
\begin{equation}
\label{su2-l} [J_{a},\ell]=[J_{a},\tilde\ell]=0 \,.
\end{equation}
Besides, the operators \p{Ta-def} commute with generators of
six-dimensional translations and with the Lorentz algebra
$\mathfrak{so}(1,5)$ generators \p{PM-gen}, \p{M-1}:
\begin{equation}
\label{M-J} [P_{\mu},J_{a}]=0\,,\qquad [M_{\mu\nu},J_{a}]=
[S_{\mu\nu},J_{a}]=0\,.
\end{equation}
As a result, the $6D$ irreducible massless representations of
infinite spin are characterized by two quantum numbers: fixed real
number $\mu \in \mathbb{R}_{>0}$ and fixed (half-)integer number $s \in \mathbb{Z}_{\geq 0}/2$.

In the following, we will consider two infinite spin
representations: ``coordinate'' and ``momentum''.

\subsection{``Coordinate'' representation}
In this representation, the operators $\xi^k_\alpha$ are diagonal and are
realized by multiplication by $\xi_\alpha^k$ and the operators
$\rho_k^\alpha$ are realized as derivatives
$\rho_k^\alpha=-i\dfrac{\partial}{\partial \xi_\alpha^k}.$

We solve  the second equation in \p{C4-constr} as
\begin{equation}\label{field-coord-1}
\Psi=\delta\bigl((\xi\tilde{p}\xi)-2\mu\bigr) \;\Phi(\xi)
\end{equation}
where $\Phi(\xi)$ can be decomposed in a series of $\xi$'s
\begin{eqnarray}\label{field-coord-2}
\Phi
&=&
\sum\limits_{n_1, n_2} \phi^{\alpha(n_1)|\beta(n_2)}\;\xi^1_{\alpha(n_1)}\;\xi^2_{\beta(n_2)}
\,,
\end{eqnarray}
where we use concise notation $\alpha(n):=(\alpha_{1}\ldots\alpha_{n})$ for symmetrization of $n$ spinor indices,
and the vertical line $|$ in $\phi^{\alpha(n_1)|\beta(n_2)}$
distinguishes the indices which contract with the indices of the spinors $\xi^1$ and spinors $\xi^2$.
Also we use the notation $\xi_{\alpha(n)}:=\xi_{\alpha_1}\ldots\xi_{\alpha_n}$ for $n$ commuting spinors $\xi$.

From the second equation in \eqref{C4-constr} we find restrictions on
the coefficients $\phi^{\alpha(n_1)|\beta(n_2)}$ in the form
\begin{eqnarray}
\label{coord-trans}
&&
n_1n_2\;p_{\alpha\beta}\;\phi^{\alpha(n_1)|\beta(n_2)}
=\mu\;\phi^{\alpha(n_1-1)|\beta(n_2-1)}\,.
\end{eqnarray}

Let us turn to the sixth-order Casimir operator $C_6$. Due to the relation
\begin{equation}
[(\xi^i\rho^{j}),(\xi\tilde p\xi)]
=
i\varepsilon^{ij}(\xi\tilde{p}\xi)
\end{equation}
we have
\begin{eqnarray}
C_6\Psi&=&\delta\bigl((\xi\tilde{p}\xi)-2\mu\bigr) \;C_6\;\Phi(\xi)\,,
\\ [6pt]
C_6\Phi(\xi)
&=&
-\frac{\mu^2}{4}
\Bigl[
(N_{\xi^1}-N_{\xi^2})^2
-2(\xi^1\rho_2)(\xi^2\rho_1)
-2(\xi^2\rho_1)(\xi^1\rho_2)
\Bigr]\Phi(\xi)\,,
\end{eqnarray}
where
\begin{equation}\label{N-xi-expr}
N_{\xi^1}=i\xi^1\rho_1\,,
\qquad
N_{\xi^2}=i\xi^2\rho_2\,,
\end{equation}
and we use here index-free expressions: $\xi^1\rho_1:=\xi^1_\alpha\rho^\alpha_1$ and etc.

Now there are two possibilities to make $\Phi(\xi)$ an eigenfunction of the sixth-order Casimir
operator $C_6$
\begin{eqnarray}
\label{xi1rho2}
\mbox{\bf a)}\quad(\xi^1\rho_2)\;\Phi(\xi)=0
\qquad\qquad \text{or} \qquad\qquad
\mbox{\bf b)}\quad(\xi^2\rho_1)\;\Phi(\xi)=0 \,.
\end{eqnarray}
We consider the first one\footnote{The second condition (\ref{xi1rho2}b) gives a
similar result when the indices 1 and 2 are
interchanged.}  (\ref{xi1rho2}a).
This condition leads to the vanishing of all terms in the expansion \p{field-coord-2},
except for the terms which have the form
\begin{equation}\label{mon-xi-nozero}
\phi^{\alpha_1\ldots\alpha_{n_1}|\beta_1\ldots\beta_{n_2}}\;
\xi^1_{[\alpha_1}\xi^2_{\beta_1]}\ldots\xi^1_{[\alpha_{n_2}}\xi^2_{\beta_{n_2}]}
\xi^1_{\alpha_{n_2 +1}}\ldots\xi^1_{\alpha_{n_1}}\,,\qquad n_1\geq n_2\,.
\end{equation}
On the function \p{mon-xi-nozero}, the operators $N_{\xi1}$ and $N_{\xi2}$, defined in \p{N-xi-expr},
take the values $n_1$ an $n_2$, respectively.
The coefficient $\phi^{\alpha(n_1)|\beta(n_2)}$ in \p{mon-xi-nozero} obeys the symmetry condition
\begin{eqnarray}
\label{coord-ir}
\phi^{(\alpha_1\ldots\alpha_{n_1}|\beta_1)\beta_2\ldots\beta_{n_2}}=0\,,
\end{eqnarray}
where $|$ distinguishes, as we mentioned above, the indices of $\xi^1$ and $\xi^2$,
and the brackets $(\ldots)$ denote symmetrization.
Thus, this tensor $\phi^{\alpha(n_1)|\beta(n_2)}$
corresponds to the two-row
Young diagram  $\mathbf{Y}(n_1,n_2)$:
$$
\underbrace{\overbrace{{\scriptsize
\ \begin{array}{|c|c|c|c|c|c|}
\hline
\; & \! \dots \! & \; & \; & \! \dots \! & \; \\ [0.15cm]
\hline
\; & \! \dots \! & \; & \multicolumn{3}{c}{} \\ [0.15cm]
\cline{1-3}
\end{array} \
}}^{n_1}\!\!\!\!\!\!\!\!\!\!\!\!\!\!\!\!\!\!\!\!\!}_{n_2}
$$
If the condition (\ref{xi1rho2}a) is satisfied, then using
\begin{eqnarray}
[(\xi^2\rho_1),(\xi^1\rho_2)]
=
N_{\xi^1}-N_{\xi^2}\,,
\end{eqnarray}
we find that
\begin{eqnarray}
C_6\Phi(\xi)
&=&
-\mu^2\;s(s+1)\;\Phi(\xi)\,,
\qquad
s=(n_1-n_2)/2\,,
\end{eqnarray}
and the eigenfunctions of the sixth-order Casimir operator $C_6$ are
\begin{eqnarray}
\label{coord-Phi_s}
\Phi_s(\xi)&=&\sum_{n=0}^\infty \phi^{\alpha(2s+n)|\beta(n)}\;\xi^1_{\alpha(2s+n)}\;\xi^2_{\beta(n)}
\,,
\end{eqnarray}
with fixed (half)integer value $s \in \mathbb{Z}_{\geq 0}/2$. Thus, we get a
description of the infinite spin field in terms of
the field \eqref{coord-Phi_s} which must satisfy the massless condition
\eqref{P2-0}, transversality condition
\eqref{coord-trans} and irreducibility with respect to index
permutations \eqref{coord-ir}:
\begin{eqnarray}
\label{m=0}
&&p^2\,\phi^{\alpha_1\ldots\alpha_{2s+n}|\beta_{1}\ldots\beta_{n}}=0\,,
\qquad
\qquad
\qquad
n=0,1,2,\ldots\,,
\\[0.5em]
&&
\label{div}
n(2s+n)\,p_{\alpha_{2s+n}\beta_n}\,\phi^{\alpha_1\ldots\alpha_{2s+n}|\beta_{1}\ldots\beta_{n}}
=\mu\;\phi^{\alpha_1\ldots\alpha_{2s+n-1}|\beta_{1}\ldots\beta_{n-1}}\,,
\\[0.5em]
\label{young}
&&\phi^{(\alpha_1\ldots\alpha_{2s+n}|\alpha)\beta_{1}\ldots\beta_{n-1}}=0\,.
\end{eqnarray}
Namely, these conditions describe the irreducible infinite spin field
and, therefore, these conditions must be consequences of the
equations of motion appearing from the true infinite spin Lagrangian.

Let us turn to the ``momentum'' representation.

\subsection{``Momentum'' representation}

The ``momentum'' representation is similar to the ``coordinate''
representation considered before, only the roles of $\xi$'s and
$\rho$'s are interchanged. In the ``momentum'' representation, the
operators $\rho_k^\alpha$ are diagonal and are realized by
multiplication by $\rho_k^\alpha$ and the operators $\xi^k_\alpha$
are realized as derivatives $\xi^k_\alpha=i\dfrac{\partial}{\partial
\rho_k^\alpha}.$
%and the field has the following dependence $\Psi(p,\xi)$.

We begin with solution to the first equation in \eqref{C4-constr} in
the form
\begin{equation}\label{field-coord-1-b}
\tilde\Psi=\delta\bigl((\rho p\rho)-2\mu\bigr) \;\tilde\Phi(\rho),
\end{equation}
where $\Phi(\rho)$ can be decomposed in a series in $\rho$'s:
\begin{eqnarray}\label{field-coord-2-b}
\tilde\Phi
&=&
\sum\limits_{n_1,n_2} \varphi_{\alpha(n_1)|\beta(n_2)}\;\rho_1^{\alpha(n_1)}\;\rho_2^{\beta(n_2)}
\,,
\end{eqnarray}
where we use the conventions from the previous subsection
with the replacement of spinors $\xi$ with subscripts by spinors $\rho$ with superscripts.
From the second equation in \eqref{C4-constr} we get the restrictions
on the coefficients $\varphi_{\alpha(n_1)|\beta(n_2)}$
\begin{eqnarray}
\label{mom-trans}
&&
n_1n_2\;\tilde{p}^{\alpha\beta}\;\varphi_{\alpha(n_1)|\beta(n_2)}
=-\mu\;\varphi_{\alpha(n_1-1)|\beta(n_2-1)}\,.
\end{eqnarray}

Using the relation
\begin{equation}
[(\xi^i\rho^{j}),(\rho p\rho)]
=
i\varepsilon^{ij}(\rho p\rho)
\end{equation}
in  the sixth-order Casimir operator $C_6$, one gets
\begin{eqnarray}
C_6\,\tilde\Psi&=&\delta\bigl((\rho p\rho)-2\mu\bigr) \;C_6\;\tilde\Phi(\rho)\,,
\\ [6pt]
C_6\,\tilde\Phi(\rho)
&=&
-\frac{\mu^2}{4}
\Bigl[
(N_{\rho_1}-N_{\rho_2})^2
-2(\rho_2\xi^1)(\rho_1\xi^2)
-2(\rho_1\xi^2)(\rho_2\xi^1)
\Bigr]\tilde\Phi(\rho)\,,
\\ [6pt]
&&
N_{\rho_1}=-i\rho_1\xi^1\,,
\quad
N_{\rho_2}=-i\rho_2\xi^2
\,. \label{N-rho-expr}
\end{eqnarray}
Again, there are two possibilities to make $\tilde\Phi(\rho)$
an eigenfunction of the sixth-order Casimir operator $C_6$
\begin{eqnarray}
\label{rho1xi2}
\mbox{\bf a})\quad(\rho_1\xi^2)\tilde\Phi(\rho)=0
\qquad\qquad
\text{or}
\qquad\qquad
\mbox{\bf b)}\quad(\rho_2\xi^1)\tilde\Phi(\rho)=0
\,.
\end{eqnarray}
We consider the first one without loss of generality.
As a solution to condition (\ref{rho1xi2}a), we get the field \p{field-coord-2-b} containing only terms of the following form:
\begin{equation}\label{mon-rho-nozero}
\varphi_{\alpha_1\ldots\alpha_{n_1}|\beta_1\ldots\beta_{n_2}}\;
\rho_1^{[\alpha_1}\rho_2^{\beta_1]}\ldots\rho_1^{[\alpha_{n_2}}\rho_2^{\beta_{n_2}]}
\rho_1^{\alpha_{n_2 +1}}\ldots\rho_1^{\alpha_{n_1}}\,,\qquad n_1\geq n_2\,.
\end{equation}
All other terms in the expansion \p{field-coord-2-b} are equal to zero.
The coefficient $\varphi_{\alpha(n_1)|\beta(n_2)}$ in \p{mon-rho-nozero} satisfies the symmetry condition
\begin{eqnarray}
\label{mom-ir}
\varphi_{(\alpha_1\ldots\alpha_{n_1}|\beta_1)\beta_2\ldots\beta_{n_2}}=0
\end{eqnarray}
and corresponds to the two-row Young diagram
$\mathbf{Y}(n_1,n_2)$:
$$
\underbrace{\overbrace{{\scriptsize
\ \begin{array}{|c|c|c|c|c|c|}
\hline
\; & \! \dots \! & \; & \; & \! \dots \! & \; \\ [0.15cm]
\hline
\; & \! \dots \! & \; & \multicolumn{3}{c}{} \\ [0.15cm]
\cline{1-3}
\end{array} \
}}^{n_1}\!\!\!\!\!\!\!\!\!\!\!\!\!\!\!\!\!\!\!\!\!}_{n_2}
$$
The operators $N_{\rho1}$ and $N_{\rho2}$, defined in \p{N-rho-expr},
take the values $n_1$ and $n_2$ on the function \p{mon-rho-nozero}, respectively
and satisfy the equality $[(\rho_2\xi^1),(\rho_1\xi^2)] =N_{\rho_1}-N_{\rho_2}$.
Therefore, if the condition (\ref{rho1xi2}a) is true,
one gets that
\begin{eqnarray}
C_6\,\tilde\Phi(\rho)
&=&
-\mu^2\;s(s+1)\;\tilde\Phi(\rho)\,,
\qquad
s= (n_1-n_2)/2\,,
\end{eqnarray}
and the eigenfunctions of the sixth-order Casimir operator $C_6$ are
\begin{eqnarray}
\label{mom-Phi_s}
\tilde\Phi_s(\rho)&=&\sum_{n=0}^\infty \varphi_{\alpha(2s+n)\beta(n)}\;\rho_1^{\alpha(2s+n)}\;\rho_2^{\beta(n)}
\,,
\end{eqnarray}
with a fixed (half)integer value $s \in \mathbb{Z}_{\geq 0}/2$. As result, we obtain
a different description of the infinite spin representation in terms of the field \eqref{mom-Phi_s}.
This field is described by the following set of equations:
\begin{eqnarray}
\label{m=0-b}
&&p^2\,\varphi_{\alpha_1\ldots\alpha_{2s+n}|\beta_{1}\ldots\beta_{n}}=0\,,\qquad n \in \mathbb{Z}_{\geq 0}
\\[0.5em]
&&
\label{div-b}
n(2s+n)\,\tilde{p}^{\alpha_{2s+n}\beta_n}\,\varphi_{\alpha_1\ldots\alpha_{2s+n}|\beta_{1}\ldots\beta_{n}}
=
-\mu\;\varphi_{\alpha_1\ldots\alpha_{2s+n-1}|\beta_{1}\ldots\beta_{n-1}}\,,
\\[0.5em]
\label{young-b}
&&\varphi_{(\alpha_1\ldots\alpha_{2s+n}|\alpha)\beta_{1}\ldots\beta_{n-1}}=0\,,
\end{eqnarray}
which contain the massless condition, transversality
condition and irreducibility with respect to index
permutations.
These conditions also must be consequences of the equations of
motion for a true infinite spin Lagrangian.

In Sect.\,3, we present the Lagrangian construction of these infinite spin fields in the case of integer helicities.

\subsection{Covariant description in terms of $\mathrm{SU}(2)$ harmonics}

In the previous subsections, when deriving the conditions for the
irreducible infinite spin field, we singled out the components
$i=1,2$ of the $\mathrm{SU}(2)$ Majorana-Weyl spinors \p{op-s}
separately. Although these intermediate considerations were
$\mathrm{SU}(2)$-non-covariant, the final expressions for the field
and the corresponding irreducibility conditions are
$\mathrm{SU}(2)$-covariant. A covariant description can be achieved
using the $\mathrm{SU}(2)$ harmonic procedure that preserves covariance
at all stages.

For this reason, analogously to \cite{GIKOS,GIOS}, we use the commuting
variables $u_i^\pm$, which are the elements of the $\mathrm{SU}(2)$
groups, and parameterize the compact space $\mathbb{S}^2$. These two
variables are subject to the conditions $u^i{}^+ u_i^- =1$, $u_i^+
u_j^- - u_j^+ u_i^-=\epsilon_{ij}$ and the reality conditions
$(u_i^\pm)^* =\mp u^i{}^\mp$, where
$u^i{}^\pm=\epsilon^{ij}u_j^\pm$. The harmonic covariant derivatives
used have the following form \cite{GIKOS,GIOS}:
\begin{equation}
\label{D-u}
D^{\pm\pm}:= u_i^\pm \frac{\partial}{\partial u_i^\mp}\,,\qquad
D^{0}:=u_i^+ \frac{\partial}{\partial u_i^+} - u_i^- \frac{\partial}{\partial u_i^-}\,.
\end{equation}
Imposing conditions on the field (the state vector) with the use of operators \p{D-u} leads to fixing the dependence
of the field on the harmonic variables $u_i^\pm$.
There are two main cases:
\begin{description}
\item[1)]
In the first case, the conditions
\begin{equation}
\label{D-ind}
D^{++}F^{(q)}(u)=0\,,\qquad
D^{0}F^{(q)}(u)=q\,F^{(q)}(u)\,,\qquad q\geq 0
\end{equation}
have a general solution
\begin{equation}
\label{sol-ind}
F^{(q)}(u)=u_{i_1}^+\ldots u_{i_q}^+ f^{i_1 \ldots i_q}\,,
\end{equation}
where  $f^{i_1 \ldots i_q}$ do not depend on the harmonics $u_i^\pm$.
\item[2)]
In the second case, the conditions
\begin{equation}
\label{D-ind--}
D^{--}\tilde F^{(-q)}(u)=0\,,\qquad
D^{0}\tilde F^{(-q)}(u)=-q\,\tilde F^{(-q)}(u)\,,\qquad q\geq 0
\end{equation}
have general a solution
\begin{equation}
\label{sol-ind--}
\tilde F^{(-q)}(u)=u_{i_1}^-\ldots u_{i_q}^- \tilde f^{i_1 \ldots i_q}\,,
\end{equation}
where  $\tilde f^{i_1 \ldots i_q}$ do not depend on the harmonics $u_i^\pm$.
\end{description}

Thus, we expand the space of variables \p{op-v}, \p{op-s} by adding
harmonics. That is, we consider the space with the following
variables: $x^{\mu}$, $p_{\mu}$; $\xi_\alpha^i$, $\rho^{\alpha i}$;
$u_i^\pm$. Now we can move from these variables to an equivalent set of
new variables $x^{\mu}$, $p_{\mu}$; $\xi_\alpha^\pm$,
$\rho^{\alpha}{}^\pm$; $u_i^\pm$ where $\xi_\alpha^\pm:=\xi_\alpha^i
u_i^\pm$, $\rho^{\alpha}{}^\pm:=\rho^{\alpha}{}^i u_i^\pm$ are
the $\mathrm{SU}(2)$-scalars and have the following non-vanishing
commutators
\begin{equation}
\label{op-sp-com}
\left[\xi_\alpha^\pm,\rho^{\beta}{}^\mp\right]=\pm\, i\, \delta_\alpha^{\beta}\,.
\end{equation}
In the new variables
the harmonic covariant derivatives \p{D-u} take the form:
\begin{equation}
\label{D-u-n}
\mathcal{D}^{\pm\pm}=D^{\pm\pm}+J^{\pm\pm}\,,\qquad
\mathcal{D}^{0}=D^{0}+J^{0}\,,
\end{equation}
where
\begin{equation}
\label{J-n}
J^{\pm\pm}=\mp\, i\, \xi_\alpha^\pm \rho^\alpha{}^\pm\,,\qquad
J^{0}=i \left(\xi_\alpha^+ \rho^\alpha{}^- +\xi_\alpha^- \rho^\alpha{}^+\right)\,.
\end{equation}
The generators \p{J-n} form the $\mathfrak{su}(2)$ algebra
$\left[J^{++},J^{--}\right]=J^{0}$, $\left[J^{0},J^{\pm\pm}\right]=\pm \, 2\,J^{\pm\pm}$.
In terms of these operators the $\mathfrak{su}(2)$ Casimir operators $J_{a}J_{a}$
presented in \p{C6-Cas} takes the form
$$
J_{a}J_{a} \ =  \
J^{--} J^{++} + \frac12 \,J^{0}\Big(\frac12 \,J^{0} +1\Big)
\ = \
J^{++} J^{--} - \frac12 \,J^{0}\Big(-\frac12 \,J^{0} +1\Big)\,.
$$
Besides, in the new variables the operators \p{l-expr} take the form
\begin{equation}
\label{l-expr-n}
\ell \ = \ \rho^\alpha{}^- p_{\alpha\beta}\rho^\beta{}^+\,,\qquad\quad
\tilde\ell \ = \ \xi_\alpha^+ \tilde p^{\,\alpha\beta}\xi_\beta^- \,.
\end{equation}

To define the irreducible infinite spin representation, we require
the fulfillment of the field masslessness condition \p{P2-0} and the
conditions \p{C4-constr} to fix the fourth-order Casimir operator
$C_4$ where $\ell$, $\tilde\ell$ are defined in \p{l-expr-n}.

To fix the sixth-order Casimir operator \p{C6-Cas}, we study two cases.

In the first case, we consider the infinite spin field as the highest
weight vector in the representation space of the $\mathfrak{su}(2)$ algebra
\p{J-n}. It means that this field obeys the equations
\begin{equation}
\label{eq-3-a}
J^{++}\, \Psi = 0 \, ,
\qquad \big(J^0-2s\big)\, \Psi = 0 \, ,
\end{equation}
where $J^{++}$, $J^0$ are defined in \p{J-n}.

It is important to point that the use of harmonics requires the
imposition of additional harmonic constrains on the infinite spin
field. We postulate the following harmonic equations
\begin{equation}
\label{eq-5}
\mathcal{D}^{++}\, \Psi = 0 \, ,
\qquad
\big(\mathcal{D}^0-2s\big)\, \Psi = 0 \, ,
\end{equation}
where the ``longer'' (in the new basis) harmonic covariant derivatives $\mathcal{D}^{++}$, $\mathcal{D}^0$ are defined in \p{D-u-n}.
Taking into account  expressions \p{D-u-n} and equations \p{eq-3-a},
we see that equations \p{eq-5} are equivalent to the simple equations
\begin{equation}
\label{eq-7}
{D}^{++}\, \Psi = 0 \, ,
\qquad
{D}^0\, \Psi = 0 \, ,
\end{equation}
where ${D}^{++}$, ${D}^0$ are defined in \p{D-u}. However, due to
\p{D-ind} and \p{sol-ind}, these equations imply that the field $\Psi$
in the harmonic basis does not depend on the harmonic variables (in
the new basis).

Considering the representation, where the operators $\xi_\alpha^\pm$
are diagonal and realized by multiplication by $\xi_\alpha^\pm$ and
the operators $\rho^{\pm\alpha}$ are realized as $\displaystyle
\rho^{\pm\alpha} = \pm \,i\,\frac{\partial}{\partial \xi_\alpha^\mp}
$, we get that solution of the equations \p{P2-0},
\p{C4-constr}, \p{eq-3-a}, \p{eq-7} is the infinite spin field
\p{field-coord-1}, \p{field-coord-2} in which $\xi_\alpha^+$ and
$\xi_\alpha^-$ play the role of $\xi_\alpha^1$ and $\xi_\alpha^2$,
respectively. Moreover, the equations for the component fields in
the expansion \p{field-coord-2} are exactly the same as
equations \p{m=0}, \p{div}, \p{young}.

In the second case, we consider the infinite spin field as the
lowest weight vector in the representation space of the $\mathfrak{su}(2)$ generators \p{J-n}. That is, in addition to the
masslessness condition \p{P2-0} and the conditions \p{C4-constr},
the field obeys the equations
\begin{equation}
\label{eq-3-b}
J^{--}\, \tilde\Psi = 0 \, ,
\qquad
\big(J^0+2s\big)\, \tilde\Psi = 0 \, ,
\end{equation}
where $J^{--}$, $J^0$ are defined in \p{J-n}.
As additional harmonic constrains, we take the conditions
\begin{equation}
\label{eq-5-b}
\mathcal{D}^{--}\, \tilde\Psi = 0 \, ,
\qquad
\big(\mathcal{D}^0+2s\big)\, \tilde\Psi = 0 \, ,
\end{equation}
where $\mathcal{D}^{++}$, $\mathcal{D}^0$ are defined in \p{D-u-n}.
However, taking into account  expressions \p{D-u-n} and equations
\p{eq-3-b}, we see that  equations \p{eq-5-b} are equivalent to
the simple equations
\begin{equation}
\label{eq-7-b}
{D}^{--}\, \tilde\Psi = 0 \, ,
\qquad
{D}^0\, \tilde\Psi = 0 \, ,
\end{equation}
where ${D}^{--}$, ${D}^0$ are defined in \p{D-u}. As shown in
\p{D-ind--} and \p{sol-ind--}, equations \p{eq-7-b} imply that the
field  $\tilde\Psi$ in the harmonic basis does not depend on the
harmonic variables. As result, in the representation with the diagonal
realization of the operators $\rho^{\pm\alpha}$ and the realization
of the operators $\xi_\alpha^\pm$ by $\displaystyle \xi_\alpha^\pm =
\pm \,i\,\frac{\partial}{\partial \rho^{\mp\alpha}} $, we obtain
that the solution of equations \p{P2-0}, \p{C4-constr},
\p{eq-3-b}, \p{eq-7-b} is the infinite spin field
\p{field-coord-1-b}, \p{field-coord-2-b} in which $\rho^{\alpha-}$
and $\rho^{\alpha+}$ play the role of $\rho_1^{\alpha}$ and
$\rho_2^{\alpha}$, respectively. Wherein, the component fields in
expansion \p{field-coord-2-b} satisfy exactly equations
\p{m=0-b}, \p{div-b}, \p{young-b}.

%\newpage
\section{Lagrangian construction}

In this section we will describe an approach to the Lagrangian
formulation for the $6D$ infinite spin field theory in  the case of integer helicities case when $s$ is integer. The formulation
is derived within the framework of the BRST approach and is related
to using the BRST charge on the Fock space.

\subsection{Generalized Fock space}

The first step in constructing the BRST Lagrangian is to rewrite the
conditions \eqref{m=0}, \eqref{div}, \eqref{young} and
\eqref{m=0-b}, \eqref{div-b}, \eqref {young-b}, which define the
field realization of the  $6D$ irreducible infinite spin
representation as equations in the Fock space. Namely, these
equations should be reproduced as consequences of the Lagrangian
equations of motion. It is these equations that must be reproduced
as consequences of the Lagrangian equations of motion, which
essentially means a dynamic description of the considered
irreducible representation.

Let us introduce the creation $a^\dagger_\alpha$, $b^\dagger_\alpha$
and annihilation $a^\alpha$, $b^\alpha$ operators, which commute with the generators of the Heisenberg algebra \p{op-v},
to satisfy the following commutation relations:
\begin{equation}
\label{ab-com}
[a^\alpha,a^\dagger_\beta]=\delta^\alpha_\beta\,,\qquad [b^\alpha,b^\dagger_\beta]=\delta^\alpha_\beta
\end{equation}
and define the ``vacuum'' state
\begin{equation}\label{vac}
|0\rangle\,,
%\qquad \langle0|=(|0\rangle)^\dagger\,,
\qquad
\langle0|0\rangle=1
\end{equation}
by the relations
\begin{equation}
\label{caoperators}
a^\alpha|0\rangle=b^\alpha|0\rangle=0\,,
\qquad\langle0|a^\dagger_\alpha=\langle0|b^\dagger_\alpha=0
\,.
\end{equation}

Now one defines as usual the ket- and bra-vector states in the Fock
space
\begin{equation}
\label{F-sol-ser-}
|\phi_s\rangle
\ = \
\sum_{n=0}^\infty
a^\dagger_{\alpha_1}\ldots a^\dagger_{\alpha_{2s+n}} \,
b^\dagger_{\beta_1}\ldots b^\dagger_{\beta_{n}} \,
\,\phi^{\alpha_1\ldots\alpha_{2s+n}|\beta_{1}\ldots\beta_{n}}|0\rangle
\,.
\end{equation}
\begin{equation}
\label{F-sol-ser-2}
\langle\varphi_s|
\ = \
\sum_{n=0}^\infty
\langle0|
a^{\alpha_1}\ldots a^{\alpha_{2s+n}} \,
b^{\beta_1}\ldots b^{\beta_{n}} \,
\,\varphi_{\alpha_1\ldots\alpha_{2s+n}|\beta_{1}\ldots\beta_{n}}
\end{equation}
and also introduces the following operators:
\begin{eqnarray}
\label{L0}
L_0 &:=& l_0=p^\mu p_\mu\,, \\ [6pt]
L &:=& l-\mu
=
a^\alpha p_{\alpha\beta}b^\beta-\mu
\,,
\label{L}
\\ [6pt]
\tilde L &:=& \tilde l-\mu
=
b_\alpha^+\tilde{p}^{\alpha\beta}a_\beta^+-\mu
\,,
\label{tL}
\\ [6pt]
Y&:=&a^\dag_\alpha b^\alpha
\,,
\label{Y}
\\ [6pt]
\tilde Y&:=&b^\dag_\alpha a^\alpha
\,.
\label{tY}
\end{eqnarray}
The only nonzero commutator of the above operators is
\begin{equation}\label{algebra}
[\tilde L,L]=-K\,L_0 \,,
\end{equation}
where
\begin{equation}
\label{KKK}
K:=a_\alpha^\dag a^\alpha+b_\alpha^\dag b^\alpha+4
\end{equation}
has the following commutators with the operators \p{L0}-\p{tY}
\begin{equation}
\label{add1}
[K,L]=-2\,L \,, \;\;\;\;
[K,\tilde L]=2 \, \tilde L \,, \;\;\;\;
[K,L_0]= 0 \,, \;\;\;\;
[K,Y]= 0 \,, \;\;\;\;
 [K, \tilde Y]= 0\, .
\end{equation}
As a result, the conditions \eqref{m=0}, \eqref{div}, \eqref{young}
take  the form of the ket-Fock space equations:
\begin{equation}
\label{ket-eqs}
L_0|\phi_s\rangle=0\,,
\qquad
L|\phi_s\rangle=0\,,
\qquad
Y|\phi_s\rangle=0
\end{equation}
and the conditions \eqref{m=0-b}, \eqref{div-b}, \eqref{young-b}
take a form of the bra-Fock space equations:
\begin{equation}
\label{bra-eqs}
\langle\varphi_s|L_0 = 0 \,,
\qquad
\langle\varphi_s|\tilde{L} = 0\,,
\qquad
\langle\varphi_s|\tilde Y = 0\,.
\end{equation}
Thus, all the conditions defining the $6D$ infinite spin irreducible
representation are completely formulated in the Fock space.

In what follows we will consider only
 the operators $L_0$, $L$ and $\tilde{L}$ as the first class
 constraints and
construct a BRST charge in accordance with this assumption.
Additional constraints stipulated by the operators $Y$ and $\tilde Y$,
 which are central elements for the algebra
 with the generators (\ref{L0}) -- (\ref{tL}),
 (\ref{KKK}), will be put
off-shell on the states vectors.

Note that the vectors \p{F-sol-ser-}, \p{F-sol-ser-2}
are the eigenvectors of the operator
\begin{equation}
\label{Y0}
Y_0=\frac12\left(a_\alpha^\dag a^\alpha-b_\alpha^\dag b^\alpha\right)
\end{equation}
with the eigenvalue $s \in \mathbb{Z}_{\geq 0}/2$:
\begin{equation}
\label{eq-Y0}
Y_0|\phi_s\rangle \ = \ s\,|\phi_s\rangle\,,
\qquad
\langle\varphi_s|Y_0  \ = \  s\,\langle\varphi_s|\,.
\end{equation}
In fact, the consideration of states \p{F-sol-ser-}, \p{F-sol-ser-2} is the resolution of equations \p{eq-Y0}
fixing one of the quantum numbers that determine the irreducible infinite spin representation.

\subsection{BRST charge}

Consider the operators $F_a$ = ($L_0$, $L$, $\tilde L$),
defined in
\p{L0}-\p{tL} as the
first class constraints in some as yet unknown
gauge Lagrangian theory. Since these operators form a closed
algebra, $[F_a,F_b\}=f_{ab}{}^c F_c$, with the
functions $f_{ab}{}^c$ dependent on $K$
\eqref{algebra}, one can try to
build {\it the first rank BRST charge} in the standard form
\begin{equation}\label{Q-gen}
Q=c^aF_a+\frac12\,f_{ab}{}^c c^ac^b\mathcal{P}_c\,,
\end{equation}
where $c^a$ and $\mathcal{P}_a$ are the odd ghosts and their momenta (antighosts).
Indeed, by using the operators $F_a = (L_0,L,\tilde L$)
 and the corresponding
ghosts $c_a = (\eta_0, \eta^+, \eta)$, we construct
the BRST charge in the form
\begin{equation}\label{Q}
Q= \eta_0L_0+\eta^+L +\eta \tilde L
-\eta^+\eta K \mathcal{P}_0
\,,
\end{equation}
which is nilpotent\footnote{Note that the operator $K$ in relation
\eqref{algebra} does not commute with constraints
(see (\ref{add1}))
that may raise some questions when constructing
a nilpotent BRST charge. However, these commutators appear
in calculating the square $Q^2$ with the
factors $\eta^2\equiv 0$, ${\eta^+}^2\equiv 0$,
which provides nilpotency. Note also that, the algebra
\eqref{algebra} lies completely within the framework of the general
BFV construction (see, e.g. \cite{BFV1,BFV2,BFV3}), which in our case
allows us to write the BRST charge in the form \eqref{Q}.
Since the algebra with generators  \p{L0}, \p{L}, \p{tL}, \p{KKK}
is a quadratic algebra, it would be interesting
to consider the derivation of the BRST charge \p{Q}
by the methods of papers \cite{IKO1,IKO2}.}:
\begin{equation}\label{Q2}
Q^2 =0\,.
\end{equation}
In addition, the BRST charge \p{Q} commutates with the constraint \p{Y}:
\begin{equation}\label{QY}
[Q,Y]=[Q,\tilde Y]=0\,.
\end{equation}

The BRST-charge (\ref{Q}) acts in the extended Fock space where the ghost operators obey the (anti)commutation relations
\begin{equation}\label{ghosts}
\{\eta,{\cal{}P}^+\}
= \{{\cal{}P}, \eta^+\}
=\{\eta_0,{\cal{}P}_0\}
=1
\end{equation}
and act on the ``vacuum'' vector as follows:
\begin{equation}
\eta|0\rangle=\mathcal{P}|0\rangle=\mathcal{P}_0|0\rangle
=0\,.
\end{equation}
They possess the standard  ghost numbers,
$gh(\mbox{``coordinates''}) =  - gh(\mbox{``momenta''}) = 1$,
providing the property  $gh(Q) = 1$.
The ket-vectors and bra-vectors of the extended Fock space are decomposed into ghost variables as follows:
\begin{eqnarray}\label{ext-vector}
|\Phi_s\rangle
&=&
|\phi_s\rangle+\eta^+\mathcal{P}^+|\phi_{s2}\rangle +\eta_0\mathcal{P}^+|\phi_{s3}\rangle
\,,
\\ [6pt]
\langle\tilde\Phi_s|
&=&
\langle\varphi_s|+\langle\varphi_{s2}|\mathcal{P}\eta+\langle\varphi_{s3}|\mathcal{P}\eta_0
 \,,
 \label{ext-vector2}
\end{eqnarray}
where $|\phi_s\rangle$, $|\phi_{s2}\rangle$, $|\phi_{s3}\rangle$ have the form \p{F-sol-ser-},
whereas $\langle\varphi_s|$, $\langle\varphi_{s2}|$, $\langle\varphi_{s3}|$ are represented as \p{F-sol-ser-2}.

Thus, the constraints \p{L0}, \p{L}, \p{tL} are taken into account by the BRST charge \p{Q}, which is built on them.

The remaining constraints (last in \p{ket-eqs} and \p{bra-eqs}) are taken into account by a special form of  the ket- and bra-states  $|\Phi_s\rangle$ and $\langle\tilde\Phi_s|$. Namely, due to  \eqref{QY}, we require that the generalized states \p{ext-vector}, \p{ext-vector2} obey
the conditions\footnote{A similar procedure for taking into account certain constraints
in the BRST formalism was used in \cite{AlkGr,ACG18}.}
\begin{equation}
Y|\Phi_s\rangle=0\,,
\qquad \langle\tilde\Phi_s|\tilde Y=0\,,
\label{Y-eq-ket}
\end{equation}
which lead to the following equations in the components:
\begin{equation}
Y|\phi_s\rangle=Y|\phi_{s2}\rangle=Y|\phi_{s3}\rangle=0\,,
\qquad
\langle\varphi_s|\tilde Y=\langle\varphi_{s2}|\tilde Y=\langle\varphi_{s3}|\tilde Y=0
\,.
\label{Y-eq-bra}
\end{equation}
For this reason, we require that the coefficient functions
$\phi^{\alpha_1\ldots\alpha_{2s+n}|\beta_{1}\ldots\beta_{n}}$ and $\varphi_{\alpha_1\ldots\alpha_{2s+n}|\beta_{1}\ldots\beta_{n}}$
in the expansions
\p{F-sol-ser-}, \p{F-sol-ser-2} (component fields) obey conditions \p{young} and \p{young-b}.
Due to this, the equations are satisfied  automatically.

The equation of motion of this BRST-field is postulated in the form\footnote{In what follows we consider
the Fock space of ket-vectors only. Consideration of the Fock space of bra-vectors is similar.}
\begin{equation}\label{Q-eq}
Q\; |\Phi_s\rangle=0 \,.
\end{equation}
Due to the nilpotency of the BRST charge, the field \p{ext-vector}
is defined up to the gauge transformations
\begin{equation}\label{Q-var}
|\Phi'_s\rangle \ = \ |\Phi_s\rangle +Q \; |\Lambda_s\rangle \,,
\end{equation}
where the gauge parameter $|\Lambda_s\rangle$ has (since $gh(Q) = 1$
and $gh(\mathcal{P}_1^+) = - 1$) the form
\begin{equation}
|\Lambda_s\rangle \ = \ \mathcal{P}^+|\lambda_s\rangle \,.
\label{g-parameter}
\end{equation}
Like the states $|\phi_s\rangle$, $|\phi_{s2}\rangle$, $|\phi_{s3}\rangle$ in (\ref{ext-vector}), the gauge
parameter  $|\lambda_s\rangle$ (\ref{g-parameter}) has the
decompositions as in (\ref{F-sol-ser-}) with the symmetry \p{young} of the components.

It is easy to show that the equation of motion $Q|\Phi\rangle=0$
\p{Q-eq} is equivalent to the following system of equations for the
vectors $|\phi_i\rangle$:
\begin{eqnarray}
\label{eq1}&&
l_0|\phi_s\rangle-(\tilde l -\mu)|\phi_{s3}\rangle \ = \ 0\,,
\\ [5pt]
\label{eq2}&&
l_0|\phi_{s2}\rangle-(l -\mu)|\phi_{s3}\rangle \ = \ 0\,,
\\ [5pt]
\label{eq3}&&
(l -\mu)|\phi_s\rangle-
(\tilde l -\mu)|\phi_{s2}\rangle-K|\phi_{s3}\rangle \ = \ 0\,.
\end{eqnarray}
In this case, the gauge transformations
$\delta|\Phi\rangle=Q|\Lambda\rangle$ \p{Q-var} look like
\begin{equation}\label{var-BRST}
\delta|\phi_s\rangle=(\tilde l-\mu)|\lambda_s\rangle \,,\qquad
\delta|\phi_{s2}\rangle=(l-\mu)|\lambda_s\rangle  \,,\qquad
\delta|\phi_{s3}\rangle=l_0|\lambda_s\rangle\,.
\end{equation}
As a result, the $6D$ infinite spin irreducible representation with quantum numbers $\mu$ and $s$ is
formulated as the BRST system in the Fock space.

\subsection{Construction of the Lagrangian}
Let us proceed directly to the Lagrangian construction.

In the framework of the BRST approach, the Lagrangian is constructed
in the form $\langle\tilde\Phi_s|Q|\Phi_s\rangle$. However, here we should
make an important remark. The fields $\phi^{\alpha(2s+n)|\beta(n)}$
in \eqref{F-sol-ser-} with upper indices are not complex conjugate
of the fields $\varphi_{\alpha(2s+n)|\beta(n)}$ in \eqref{F-sol-ser-2}
with lower indices, so that the expression
$\langle\tilde\Phi_s|Q|\Phi_s\rangle$ is not real. Therefore, we are
forced to treat the fields that are conjugate to the fields
$\phi^{\alpha(2s+n)|\beta(n)}$ as independent. Similarly to the fields
with lower indices\footnote{It means, in fact, that bra- and
ket-vectors are not considered as independent objects.}.

Taking into account the above remark, we construct the Lagrangian in
the form
\begin{eqnarray}
\mathcal{L}
&=&
\langle\tilde\Phi_s|Q|\Phi_s\rangle
+
c.c.
\nonumber
\\[6pt]
&=&
\langle\varphi_s|\Bigl\{l_0|\phi_s\rangle-(\tilde l -\mu)|\phi_{s3}\rangle\Bigr\}
+\langle\varphi_{s2}|\Bigl\{l_0|\phi_{s2}\rangle-(l -\mu)|\phi_{s3}\rangle\Bigr\}
\nonumber
\\ [6pt]
&&{}+
 \langle\varphi_{s3}|\Bigl\{(l -\mu)|\phi_s\rangle-
(\tilde l -\mu)|\phi_{s2}\rangle-K|\phi_{s3}\rangle\Bigr\}
\ + \ c.c.
\label{LagrFock}
\end{eqnarray}
The Lagrangian \eqref{LagrFock} is real by construction and
invariant under the gauge transformations \eqref{Q-var} or
\eqref{var-BRST}. It is easy to show by direct calculations that the
equations of motion for the Lagrangian \eqref{LagrFock}
automatically reproduce equations \p{eq1}--\p{eq3} as their
consequences.
As a result, one concludes that the Lagrangian
\eqref{LagrFock} describes the field dynamics of the $6D$ infinite spin
irreducible representation.
This conclusion is a direct consequence of the fact that the proposed BRST construction
takes into account all the necessary conditions \p{L0}, \p{L}, \p{tL}, \p{Y}
that determine these irreducible representations.
In this case, the constraints \p{L0}, \p{L}, \p{tL} are taken into account by the form of the BRST charge,
while the constraint \p{Y} is resolved by the form of states or fields used.

The next step is deriving the component form of the Lagrangian
\eqref{LagrFock}. This is carried out by a direct identical calculation
of oscillator-like vacuum matrix elements. The final result has the
form:
\begin{eqnarray}
\mathcal{L}
&=&
\sum_{n=0}^\infty (2s+n)n\;
\varphi_{\alpha(2s+n)|\beta(n)}
\Bigl\{
-\partial^2\phi^{\alpha(2s+n)|\beta(n)}
-i\partial^{\alpha\beta}\phi_3^{\alpha(2s+n-1)|\beta(n-1)}
+\mu\, \phi_3^{\alpha(2s+n)|\beta(n)}
\Bigr\}
\nonumber
\\ [6pt]
&&{}+
\sum_{n=0}^\infty (2s+n)n\;
\varphi_{2\alpha(2s+n)|\beta(n)}
\Bigl\{
-\partial^2\phi_2^{\alpha(2s+n)|\beta(n)}
\nonumber
\\ [6pt]
&&\hspace{18ex}{}
+i(2s+n+1)(n+1)\partial_{\gamma\delta}\phi_3^{\alpha(2s+n)\gamma|\delta\beta(n-1)}
+\mu\, \phi_3^{\alpha(2s+n)|\beta(n)}
\Bigr\}
\nonumber
\\ [6pt]
%%%%%%%%%%%%%%%%%%
&&{}+
\sum_{n=0}^\infty (2s+n)n\;
\varphi_{3\alpha(2s+n)|\beta(n)}
\Bigl\{
-i\partial^{\alpha\beta}\phi^{\alpha(2s+n-1)|\beta(n-1)}
\nonumber
\\ [6pt]
&&\hspace{18ex}{}
-i(2s+n+1)(n+1)\partial_{\gamma\delta}\phi_2^{\alpha(2s+n)\gamma|\delta\beta(n-1)}
-2(s+n+2)\phi_3^{\alpha(2s+n)|\beta(n)}
\nonumber
\\ [6pt]
&&\hspace{18ex}{} +\mu\, \phi_2^{\alpha(2s+n)|\beta(n)} -\mu\,
\phi^{\alpha(2s+n)|\beta(n)} \Bigr\}
%%%%%%%%%%%%%%%%%%%%%
\ + \ c.c.
%\; .
\label{Lagr-comp}
\end{eqnarray}
Similarly, one derives the component form of the gauge
transformations \eqref{var-BRST}. The result is written as follows:
\begin{eqnarray}
\delta\phi^{\alpha(2s+n)|\beta(n)}
&=&
i\partial^{\alpha\beta}\lambda^{\alpha(2s+n-1)|\beta(n-1)}
-\mu\,\lambda^{\alpha(2s+n)|\beta(n)}
\\ [6pt]
\delta\phi_2^{\alpha(2s+n)|\beta(n)}
&=&
-i(2s+n+1)(n+1)\partial_{\gamma\delta}\lambda^{\alpha(2s+n)\gamma|\delta\beta(n-1)}
-\mu\,\lambda^{\alpha(2s+n)|\beta(n)}
\\ [6pt]
\delta\phi_3^{\alpha(2s+n)|\beta(n)}
&=&
-\partial^2\lambda^{\alpha(2s+n)|\beta(n)}
\end{eqnarray}

The equations of motion, corresponding to the Lagrangian
\eqref{Lagr-comp}, automatically yield equations
\p{eq1}--\p{eq3}. The last equations are equivalent
to the conditions \eqref{m=0}, \eqref{div}, \eqref{young} which define
the $6D$ infinite spin irreducible representations. It can be done
as follows. First, we eliminate the fields $\phi_3$ using their gauge
transformations. Second, there is still a residual gauge freedom with
restricted gauge parameters $\lambda$,
$\partial^2\lambda=0$. Third, after eliminating the fields $\phi_3$,
the equations for $\phi_2$ take the form
$\partial^2\phi_2=0$ which coincide with the restrictions on the
parameters $\lambda$. Using the gauge transformations with these
restricted gauge parameters, we eliminate the fields $\phi_2$. Fourth,
after eliminating the auxiliary fields $\phi_2$ and $\phi_3$, the
equations of motion \eqref{eq1} and \eqref{eq2}, following from the
Lagrangian, take the form of the needed conditions \eqref{m=0},
\eqref{div} and the third condition \eqref{young} will be fulfilled
automatically
due to the imposition of the condition \p{Y-eq-ket} on the state vector
and to the resulting symmetry of the component fields, which is described by
the Young  tableaux from the previous section.

As a result, we conclude finally that the Lagrangian
\eqref{LagrFock} correctly describes  the field dynamics of the $6D$
infinite spin irreducible representation.

\section{Summary and outlook}
Let us briefly summarize. We have developed a procedure for deriving the
Lagrangian formulation for free fields forming the bosonic infinite
spin irreducible representations of the six-dimensional Poincar\'e
group.

First, we have presented a field description of infinite spin
irreducible representations in the six-dimensional Minkowski space. It
is shown that such a description is efficiently realized in the
space where the conventional phase with the space-time coordinates
$x^{\mu}$ and momenta $p_{\mu}$ is extended by the spinor coordinates
$\xi_\alpha^i$ and the corresponding spinor momenta $\rho^{\alpha
i}$. The generators of the six-dimensional Poincar\'e group were
derived in this phase space and the second-order, fourth-order and
sixth-order Casimir operators were constructed in the explicit form.
Using this phase space, we have introduced the ``coordinate'' and
``momentum'' representations and formulated the operator conditions
defining the $6D$ infinite spin irreducible representations. In
addition, we presented the obtained field description in the
framework of the harmonic approach when the $\mathrm{SU}(2)$
covariance is preserved at all stages.

Second, we have worked out the Lagrangian formulation of the theory
under consideration.  The conditions defining the irreducible
representation were reformulated as constraints on the vectors of the
Fock space. Using these constraints, we have built the BRST charge.
This allowed
us to find  the Lagrangian and the corresponding gauge
transformations in terms of the Fock space vectors and then rewrite the
results in terms of the conventional spin-tensor fields in the $6D$
Minkowski space. As the last essential point, we have proved that the
equations of motion for the constructed Lagrangian exactly reproduce
the conditions defining the infinite spin irreducible representation
of the six-dimensional Poincar\'e group. This finally confirms the
correctness of the constructed Lagrangian formulation.

The results obtained can be generalized in different directions.
First, we note that the resulting Lagrangian description leads
to second-order field equations for an infinite set of component fields,
which naturally implies their bosonic character and their integer helicities.
The next task is the construction of the Lagrangian formulation for fermionic
infinite spin fields and the corresponding supersymmetric formulation.
Second, it would be interesting to study the possibilities of finding
the interaction vertices for $6D$ infinite spin fields. Third,
problems related to eight and ten dimensions are completely open.
Although the algebraic description of the infinite spin representation of
the Poincar\'e group in any dimensions is known, the corresponding
field realizations, as far as we know, have not been developed. We
plan to study all these issues in the forthcoming works.

\section*{Acknowledgments}

I.L.B. is thankful to P.\, Schuster for useful discussions on
various aspects of the infinite spin field theory. The work of I.L.B
and S.A.F was supported by the Russian Science Foundation, project
No 21-12-00129, the work of V.A.K. was supported in part by the
Ministry of Education of the Russian Federation, project
QZOY-2023-0003.


\begin {thebibliography}{99}


\bibitem{BFIP}
I.L.\,Buchbinder, S.A.\,Fedoruk, A.P.\,Isaev, M.A.\,Podoinitsyn,
{\it Massless finite and infinite spin representations of
Poincar\'{e} group in six dimensions}, Phys. Lett.  {\bf B813}
(2021) 136064, {\tt arXiv:2011.14725\,[hep-th]}.



\bibitem{BFI-21}
I.L.\,Buchbinder, S.A.\,Fedoruk, A.P.\,Isaev, {\it Twistor
formulation of massless $6D$ infinite spin fields}, Nucl. Phys.
{\bf B973} (2021) 115576, {\tt arXiv:2108.04716\,[hep-th]}.


\bibitem{BFI-22}
I.L.\,Buchbinder, S.A.\,Fedoruk, A.P.\,Isaev, {\it Light-front
description of infinite spin fields in six-dimensional Minkowski
space}, Eur. Phys. J.  {\bf C82} (2022) 733, {\tt
arXiv:2207.02640\,[hep-th]}.

\bibitem{W}
S.\,Weinberg, {\it Massless Particles in Higher Dimensions}, Phys.
Rev. {\bf D 102} (2020) 095022, {\tt arXiv:2010.05823\,[hep-th]}.

\bibitem{KUZ}
S.M. Kuzenko, A.E. Pindur, {\it Massless particles in five and
higher dimensions}, Phys. Lett. {\bf B 812} (2021) {\tt
arXiv:2010.07124\,[hep-th]}.

\bibitem{Wigner39}
E.P.\,Wigner,
{\it On unitary representations of the inhomogeneous Lorentz group},
Annals Math.  {\bf 40} (1939) 149.

\bibitem{Wigner47}
E.P.\,Wigner,
{\it Relativistische Wellengleichungen},
Z. Physik  {\bf 124} (1947) 665.

\bibitem{BargWigner}
V.\,Bargmann, E.P.\,Wigner,
{\it Group theoretical discussion of relativistic wave equations},
Proc. Nat. Acad. Sci. US  {\bf 34} (1948) 211.


\bibitem{Isaev1}
A.P.\,Isaev, V.A.\,Rubakov,  {\it Theory Of Groups And Symmetries
(I): Finite Groups, Lie Groups, And Lie Algebras}, World Scientific,
2019, 476 pp.

\bibitem{Isaev2}
A.P.\,Isaev, V.A.\,Rubakov, {\it Theory of Groups and Symmetries
(II): Representations of Groups and Lie Algebras, Applications},
World Scientific, 2021, 600 pp.

\bibitem{BB1}
X. Bekaert, N. Boulanger, {\it The unitary representations of the
Poincar\'e group in any spacetime dimension,} Lectures presented at
2nd Modave Summer School in Theoretical Physics, 6-12 Aug 2006,
Belgium, {\tt arXiv:hep-th/0611263}.

\bibitem{BB2} X. Bekaert, N. Boulanger, {\it Tensor gauge fields in
arbitrary representations of $GL(D,R)$}, Commun. Math. Phys. {\bf
271} (2007), {\tt arXiv:hep-th/0606198}.


\bibitem{BekSk}
X.\,Bekaert, E.D.\,Skvortsov, {\it Elementary particles with
continuous spin}, Int. J. Mod. Phys.   {\bf A32} (2017) 1730019,
{\tt arXiv:1708.01030\,[hep-th]}.

\bibitem{BekMou}
X.\,Bekaert, J.\,Mourad, {\it The continuous spin limit of higher
spin field equations}, JHEP  {\bf 0601} (2006) 115, {\tt
arXiv:hep-th/0509092}.

\bibitem{Bekaert:2017xin}
X.\,Bekaert, J.\,Mourad, M.\,Najafizadeh, {\it Continuous-spin field
propagator and interaction with matter}, JHEP {\bf 1711} (2017) 113,
{\tt arXiv:1710.05788 [hep-th]}.

\bibitem{Najafizadeh:2017tin}
M.\,Najafizadeh, {\it Modified Wigner equations and continuous spin
gauge field}, Phys. Rev. D {\bf 97} (2018) 065009, {\tt
arXiv:1708.00827\,[hep-th]}.


\bibitem{HabZin}
M.V.\,Khabarov, Yu.M.\,Zinoviev, {\it Infinite (continuous) spin
fields in the frame-like formalism}, Nucl. Phys. {\bf B928} (2018)
182, {\tt arXiv:1711.08223\,[hep-th]}.

\bibitem{AlkGr}
K.B.\,Alkalaev, M.A.\,Grigoriev, {\it Continuous spin fields of
mixed-symmetry type}, JHEP {\bf 1803} (2018) 030, {\tt
arXiv:1712.02317\,[hep-th]}.

\bibitem{Metsaev18}
R.R.\,Metsaev, {\it BRST-BV approach to continuous-spin field},
Phys. Lett. {\bf B781} (2018) 568, {\tt arXiv:1803.08421\,[hep-th]}.

\bibitem{BFIR}
I.L.\,Buchbinder, S.\,Fedoruk, A.P.\,Isaev, A.\,Rusnak, {\it Model
of massless relativistic particle with continuous spin and its
twistorial description}, JHEP  {\bf 1807} (2018) 031, {\tt
arXiv:1805.09706\,[hep-th]}.


\bibitem{BuchKrTak}
I.L.\,Buchbinder, V.A.\,Krykhtin, H.\,Takata, {\it BRST approach to
Lagrangian construction for bosonic continuous spin field}, Phys.
Lett.   {\bf B785} (2018) 315, {\tt arXiv:1806.01640\,[hep-th]}.

\bibitem{R}
V.O.\,Rivelles,  {\it A Gauge Field Theory for Continuous Spin
Tachyons}, {\tt arXiv:1807.01812\, [hep-th]}.

\bibitem{BuchIFKr}
I.L.\,Buchbinder, S.\,Fedoruk, A.P.\,Isaev, V.A.\,Krykhtin, {\it
Towards Lagrangian construction for infinite half-integer spin
field}, Nucl. Phys. {\bf B958} (2020) 115114, {\tt
arXiv:2005.07085\,[hep-th]}.

\bibitem{ACG18}
K.\,Alkalaev, A.\,Chekmenev, M.\,Grigoriev, {\it Unified formulation
for helicity and continuous spin fermionic fields}, JHEP {\bf 1811}
(2018) 050, {\tt arXiv:1808.09385\,[hep-th]}.

\bibitem{Metsaev18a}
R.R.\,Metsaev, {\it Cubic interaction vertices for massive/massless
continuous-spin fields and arbitrary spin fields}, JHEP {\bf 1812}
(2018) 055, {\tt arXiv:1809.09075\,[hep-th]}.

\bibitem{BFI}
I.L.\,Buchbinder, S.\,Fedoruk, A.P.\,Isaev, {\it Twistorial and
space-time descriptions of massless infinite spin (super)particles
and fields}, Nucl. Phys. B {\bf 945} (2019) 114660, {\tt
arXiv:1903.07947[hep-th]}.

\bibitem{Metsaev19}
R.R.\,Metsaev, {\it Light-cone continuous-spin field in AdS space},
Phys. Lett. {\bf B793} (2019) 134; {\tt arXiv:1903.10495\,[hep-th]}.

\bibitem{BKSZ}
I.L.\,Buchbinder, M.V.\,Khabarov, T.V.\,Snegirev, Yu.M.\,Zinoviev,
{\it Lagrangian formulation for the infinite spin $N=1$
supermultiplets in $d=4$}, Nucl. Phys. {\bf B 946} (2019) 114717,
{\tt arXiv:1904.05580 [hep-th]}.

\bibitem{MN20}
M.\,Najafizadeh, {\it Supersymmetric Continuous Spin Gauge Theory},
JHEP {\bf 2003} (2020) 027, {\tt arXiv:1912.12310 [hep-th]}.

\bibitem{MN22}
M.\,Najafizadeh, {\it Off-shell Supersymmetric Continuous Spin Gauge
Theory}, JHEP {\bf 02} (2022) 038, {\tt arXiv:2112.10178 [hep-th]}.

\bibitem{BuchIFKr22}
I.L.\,Buchbinder, S.A.\,Fedoruk, A.P.\,Isaev, V.A.\,Krykhtin, {\it
On the off-shell superfield Lagrangian formulation of 4D, N=1
supersymmetric infinite spin theory}, Phys. Lett. {\bf B 829} (2022)
137139, {\tt arXiv:2203.12904 [hep-th]}.

\bibitem{STZ}
P.\,Schuster, N.\,Toro, K.\,Zhou, {\it Interactions of Particles
with "Continuous Spin" Fields}, JHEP {\bf 04} (2023) 010, {\tt
arXiv:2303.04816 [hep-th]}.

\bibitem{BKL}
I.L.\,Buchbinder, V.A.\,Krykhtin, P.M.\,Lavrov, {\it Gauge invariant
Lagrangian formulation of higher spin massive bosonic field theory
in AdS space}, Nucl. Phys. {\bf B 762} (2007) 344-376, {\tt
arXiv:hep-th/0608005}.

\bibitem{BKS}
I.L.\,Buchbinder, V.A.\,Krykhtin, T.V.\,Snegirev, {\it Cubic
interactions of D4 irreducible massless higher spin fields within
BRST approach}, EPJ {\bf C 82} (2022) 10071, {\tt arXiv:2208.04409
[hep-th]}.

\bibitem{GIKOS}
A.\,Galperin, E.\,Ivanov, S.\,Kalitsyn, V.\,Ogievetsky, E.\,Sokatchev,
{\it Unconstrained N=2 Matter, Yang-Mills and Supergravity Theories in Harmonic Superspac},
Class. Quant. Grav. {\bf 1} (1984) 469.

\bibitem{GIOS}
A.S.\,Galperin, E.A.\,Ivanov, V.I.\,Ogievetsky, E.S.\,Sokatchev,
{\it Harmonic Superspace},
Cambridge Univ. Press, 2001, 306 p.

\bibitem{BFV1}
I.A.\,Batalin, G.A.\,Vilkovisky {\it Relativistic S Matrix of
Dynamical Systems with Boson and Fermion Constraints}, Phys.Lett.
{\bf B 69} (1977) 309.


\bibitem{BFV2}
I.A.\,Batalin, E.S.\,Fradkin, {\it Operator Quantization of
Relativistic Dynamical Systems Subject to First Class Constraints},
Phys. Lett. {\bf B 128} (1983) 303.

\bibitem{BFV3}
I.A.\,Batalin, E.S.\,Fradkin, {\it Operatorial quantizaion of
dynamical systems subject to constraints. A Further study of the
construction}, Ann.Inst.H.Poincare Phys.Theor. {\bf 49} (1988) 145.



\bibitem{IKO1}
A.P.\,Isaev, S.O.\,Krivonos, O.V.\,Ogievetsky, {\it BRST operators for W algebras},
J. Math. Phys. {\bf 49} (2008) 073512, {\tt arXiv:0802.3781\,[math-ph]}.

\bibitem{IKO2}
A.P.\,Isaev, S.O.\,Krivonos, O.V.\,Ogievetsky, {\it BRST charges for finite nonlinear algebras},
Phys. Part. Nucl. Lett. {\bf 7} (2010) 223, {\tt arXiv:0807.1820\,[math-ph]}.



\end{thebibliography}
\end{document}